\documentclass[prb,twocolumn,a4paper,showpacs,showkeys,floatfix]{revtex4}
\usepackage{graphicx}	
\usepackage{amsmath}	
\usepackage{dcolumn}	
\renewcommand{\vec}[1]{\mbox{\boldmath$#1$}}

\begin{document}
\title{Theory and design of quantum cascade lasers in (111) $n$-type Si/SiGe}
\author{A. Valavanis}
\email{a.valavanis05@leeds.ac.uk}
\author{L. Lever}
\author{C. A. Evans}
\author{Z. Ikoni\'{c}}
\author{R. W. Kelsall}
\affiliation{Institute of Microwaves and Photonics, School of Electronic
and Electrical Engineering, University of Leeds, Leeds LS2 9JT, United
Kingdom}
\date{\today}

\begin{abstract}
  Although most work towards the realization of group IV quantum cascade
  lasers (QCLs) has focused on valence band transitions, there are many
  desirable properties associated with the conduction band. We show that
  the commonly cited shortcomings of $n$-type Si/SiGe heterostructures
  can be overcome by moving to the (111) growth direction. Specifically,
  a large band offset and low effective mass are achievable and subband
  degeneracy is preserved. We predict net gain up to lattice
  temperatures of 90\,K in a bound-to-continuum QCL with a double-metal
  waveguide, and show that a Ge interdiffusion length of at least
  8\,\AA{} across interfaces is tolerable.
\end{abstract}

\maketitle

\section{Introduction}
Quantum cascade lasers (QCLs) have been developed in a variety of III-V
materials systems, although as yet there has been no successful
demonstration in group IV\@.  A Si/SiGe QCL would potentially reduce 
fabrication costs and offer a route to photonic system-on-a-chip
applications.\cite{Pavesi2003_Kelsall}

The most common approach towards a Si/SiGe QCL uses (001) oriented 
$p$-type structures,\cite{Pavesi2003_Kelsall} and
electroluminescence\cite{ScienceDehlinger2000, PhysicaEPaul2003, 
APLLynch2002} has been demonstrated. There are however significant
challenges in designing $p$-type QCLs. The coexistence of heavy and
light holes leads to fast nonradiative scattering, and strong valence
band mixing causes large variations in transition energies and matrix
elements with in-plane wave vector.\cite{PRBIkonic2001}

The high longitudinal effective mass of SiGe $\Delta$ valleys is 
commonly regarded as a major obstacle to $n$-type 
QCLs,\cite{SSTPaul2004} and recent theoretical investigations have used
transitions in the $\Gamma$ and $L$ valleys instead.\cite{SSTHan2007,
APLDriscoll2006} We have however shown previously that a (111) oriented
Si/SiGe QCL using $\Delta$ valley transition is
viable.\cite{APLLever2008}

In this paper, we compare the strain tensors for (001) and (111)
oriented layers.  We show that (111) oriented $\Delta$ valleys offer 
larger \emph{usable\/} band offsets and lower quantization effective mass
than the (001) case and that complications due to subband degeneracy
splitting are avoided. We summarize our calculations of the principal
scattering mechanisms, current and gain, and present a
bound-to-continuum QCL design in $n$-type (111) Si/SiGe.  We calculate
the waveguide losses and predict net gain in our design.  Finally, we
investigate the effects of temperature and nonabrupt interfaces.

\section{\label{scn:SubstSel}Strain tensors}
Lattice mismatch induces strain in thin Si$_{1-x}$Ge$_x$ layers on a
relaxed Si$_{1-x_s}$Ge$_{x_s}$ substrate, where $x\neq{}x_s$. Layers
below their critical thickness deform elastically to match the in-plane
lattice constant of the substrate and strain balancing of a multilayer
structure is required to achieve mechanical stability. This is achieved
by selecting a substrate alloy which minimizes elastic potential energy
with respect to in-plane strain.\cite{Harrison2005} It is convenient to
convert between both the \emph{interface\/} coordinate system
$R=(x,y,z)$, where the $z$ axis is normal to the layer interfaces, and
the \emph{crystallographic\/} coordinate system $R'=(x',y',z')$. For the
 (001) case, $R'=R$, whereas for (111) systems a transformation matrix
$U\!\!:\!R\to{}R'$ is required.\cite{PRBHinckley1990}

The in-plane strain in $R$ is defined as
\begin{equation}
\varepsilon_{\parallel} = \frac{a_s-a}{a},
\end{equation}
where $a$ is the lattice constant of the unstrained layer and $a_s$ is 
that of the substrate.  A good approximation for lattice constant (in 
nm) is\cite{PhysStatSolBublik1974}
\begin{equation}
\label{eqn:LattConst}
a(x) = 0.5431 + 0.01992x + 0.0002733x^2.
\end{equation}

Assuming isotropy over the $xy$ plane, the strain tensors in $R'$ for 
 (001) and (111) oriented layers are\cite{SSESmirnov2004}
\begin{equation}
\varepsilon'^{(001)}=\varepsilon_{\parallel}
\begin{pmatrix}
1 & 0 & 0\\
0 & 1 & 0\\
0 & 0 & -\frac{2c'_{12}}{c'_{11}}
\end{pmatrix},
\end{equation}
\begin{equation}
\varepsilon'^{(111)}=\frac{\varepsilon_{\parallel}}{c'_{\beta}}
\begin{pmatrix}
4c'_{44}    & c'_{\alpha} & c'_{\alpha}\\
c'_{\alpha} & 4c'_{44}    & c'_{\alpha}\\
c'_{\alpha} & c'_{\alpha} & 4c'_{44}
\end{pmatrix},
\end{equation}
where $c'_{ij}$ are the elastic constants,
$c'_{\alpha}=-(c'_{11}+2c'_{12})$ and
$c'_{\beta} = {c'_{11}+2c'_{12}+4c'_{44}}$.

The minimum average strain energy corresponds to a substrate lattice 
constant,\cite{Harrison2005}
\begin{equation}
a_s=\frac{\sum\limits_k A_k l_k/a_k}{\sum\limits_k {A_k l_k}/{a_k^2}},
\end{equation}
where $l$ is the layer thickness, $k$ is the layer index and the elastic
constants are grouped into a single term $A$.  For the (001) and (111)
orientations, the elastic constants are
\begin{equation}
A_k^{(001)}=2\left(c'_{11}+c'_{12}-2\frac{c_{12}'^2}{c'_{11}}\right),
\end{equation}
\begin{equation}
A_k^{(111)}=\frac{12c'_{44}(4c'_{11}c'_{44}+8c'_{12}c'_{44}+c_{\alpha}'^2)}{c_{\beta}'^2}.
\end{equation}

\section{Band structure}
\begin{figure}[tb]
  \includegraphics{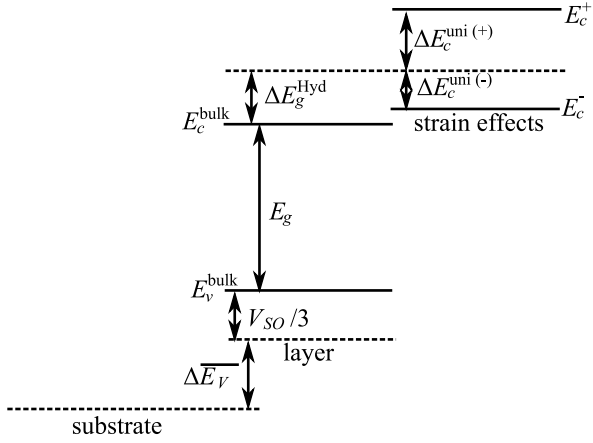}
  \caption{\label{fig:BandOffsets}Energy band schematic (not to scale)
  for a strained alloy layer on a rigid substrate. To find the
  potentials of the $\Delta$ valley minima in the strained layer, the
  difference in valence band energy and bandgap are calculated before
  adding the hydrostatic and uniaxial strain effects.}
\end{figure}

As QCLs are intersubband devices, the indirect Si$_{1-x}$Ge$_x$ 
bandgap is not an obstacle. The effective mass model for SiGe-based 
systems however, must account for transitions occurring away from the
$\Gamma$ symmetry point. The band structure is described as follows. 

\subsection{Unstrained $\Delta$ valley potential}
The unstrained conduction band minima in Si$_{1-x}$Ge$_x$ with $x<85$\%
are located in six degenerate valleys in $k$ space. Each valley lies 
close to an $X$ symmetry point, and has a spheroidal equipotential 
surface with its major axis along the associated $\Delta$
direction.\cite{Davies1998}

The relaxed $\Delta$ valley potential relative to the substrate is
determined from the model solid approximation.\cite{PRBVanDeWalle1989} 
The difference in average valence band maximum (in eV) is
\begin{equation}
  \Delta\overline{V_{\text{VB}}} = (0.74 - 0.06x_s)(x
   -x_s),\cite{PRBRieger1993}
\end{equation}
and the highest valence band maximum is one third of the spin-orbit
splitting above this, such that
\begin{equation}
  \Delta{}V_{\text{VB}} = \Delta\overline{V_{\text{VB}}} +
   \frac{\Delta{}V_{\text{SO}}}{3},
\end{equation}
where $\Delta{}V_{\text{SO}}$ is the difference in spin orbit splitting
between the two materials.\cite{PRBVanDeWalle1989}

The unstrained $\Delta$ valley conduction band offset is therefore
\begin{equation}
  \label{eqn:VCB}
  \Delta{}V_{\text{CB}} = \Delta\overline{V_{\text{VB}}} +
   \frac{\Delta{}V_{\text{SO}}}{3} + \Delta{}E_g,
\end{equation}
where $\Delta{}E_g$ is the difference in indirect $\Gamma\to\Delta$
bandgap. To a good approximation, this is given (in eV)
by\cite{PRBWeber1989}
\begin{equation}
  E_g = 1.155 - 0.43x + 0.0206x^2.
\end{equation}

\subsection{Strained layers}
\begin{figure}[tb]
  \includegraphics{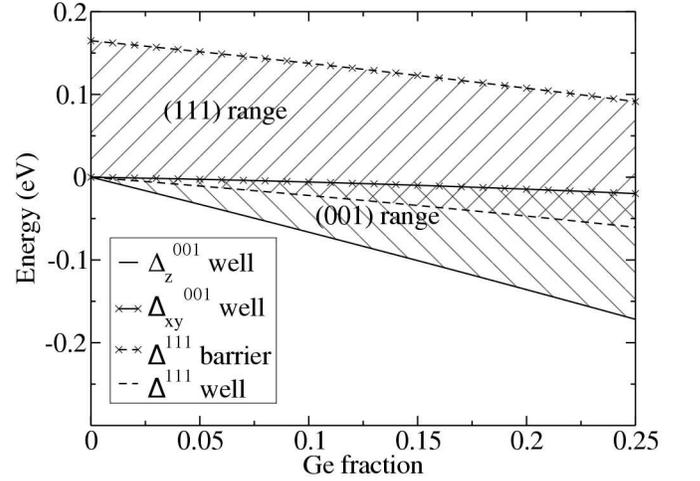}
  \caption{\label{fig:BandEnergies}Usable range of $\Delta$-valley
  offsets for (111) and (001) oriented Si wells with
  Si$_{0.5}$Ge$_{0.5}$ barriers as a function of Ge fraction in the
  substrate. For the (111) orientation the entire quantum well may be
  used whereas in (001) the $\Delta_{xy}$ well potential defines the
  upper limit.}
\end{figure}

In general, two strain dependent terms are added to Eqn.~\ref{eqn:VCB}.
Firstly, hydrostatic strain, \emph{i.e.\/}\ the overall volume change,
shifts the potential by an amount $\Delta{}V_{\text{hyd}} =
a_c\operatorname{tr}(\varepsilon')$, where $a_c$ is the hydrostatic
conduction band deformation potential. For (001) heterostructures, the
hydrostatic shift is
\begin{equation}
  \Delta{}V_{\text{hyd}}^{(001)}
    =2\left(1-\frac{c'_{12}}{c'_{11}}\right)a_c\varepsilon_{\parallel},
\end{equation}
whereas for (111) heterostructures, it is
\begin{equation}
  \Delta{}V_{\text{hyd}}^{(111)} =
   \frac{12c'_{44}}{c'_{\beta}}a_c\varepsilon_{\parallel}.
\end{equation}

Secondly, uniaxial strain, \emph{i.e.\/}\ the distortion from the cubic 
lattice form, splits the in-plane $\Delta_{xy}$ and the perpendicular
$\Delta_{z}$ valley potentials by an amount
\begin{equation}
\Delta{}V_{\text{uni}}=\left(\frac{1}{6}\pm\frac{1}{2}\right)
 \Xi_{\text{u}}(\varepsilon'_{33}-\varepsilon'_{11}),
\end{equation}
where $\Xi_{\text{u}}$ is the uniaxial deformation potential. The upper
sign represents the $\Delta_z$ valleys and the lower sign represents
$\Delta_{xy}$. This effect is absent in (111) heterostructures as strain
is identical along each of the principal crystallographic axes.

The calculation is summarized in Fig.~\ref{fig:BandOffsets}.  The 
conduction band edge is generally shifted by hydrostatic strain, and in 
the (001) orientation, uniaxial strain splits the $\Delta$ valley 
degeneracy. Although intervalley optical transitions are forbidden by 
the wave vector shift, degeneracy splitting complicates (001) QCL design
by introducing nonradiative $\Delta_z\to\Delta_{xy}$ transitions. It is
therefore desirable to restrict optical transitions to energies below
the $\Delta_{xy}$ valley minima.

For a strain balanced QCL with barriers narrower than wells, $x_s<25\%$.
Fig.~\ref{fig:BandEnergies} shows that the \emph{usable\/} energy range 
in the (001) orientation decreases almost linearly with $x_s$ from a 
maximum of 150\,meV at $x_s$=25\%.  The (111) orientation is therefore 
desirable for QCL designs as the entire $\Delta$ offset of 150\,meV may 
be used regardless of the substrate composition.

\subsection{\label{scn:EffMass}Effective mass}
\begin{table}
\caption{\label{tbl:EffMass}Quantization and average two-dimensional
density-of-states effective masses, and degeneracy $n$ of $\Delta$
valleys in (001) and (111) SiGe alloys with Ge fraction under 85\%,
using expressions derived from Ref.~\onlinecite{JAPRahman2005}.  Masses
are expressed relative to the rest mass of a free electron.}
\begin{ruledtabular}
\begin{tabular}{ccccc}
Valley                 & $m_q$                             & $m_d$                                & $n$\\
\hline
$\Delta_{xy}^{(001)}$  & $m_t$=0.19                        & $\sqrt{m_{l}m_t}$=0.42               &  4\\
$\Delta_z^{(001)}$     & $m_l$=0.916                       & $m_t$ = 0.19                         &  2\\
\hline 
$\Delta^{(111)}$       & $\frac{3m_{l}m_t}{2m_l+m_t}$=0.26 & $\sqrt{m_t\frac{2m_l+m_t}{3}}$=0.36  &  6\\
\end{tabular}
\end{ruledtabular}
\end{table}

The spheroidal $\Delta$ valley approximation remains valid in strained
SiGe, as the effective mass only varies slightly.\cite{PRBRieger1993} 
However, two separate effective masses are required: the longitudinal 
effective mass $m_l$=0.916 for wave vectors along the major axis of a
valley and the transverse effective mass $m_t$=0.19 for wave vectors
along the minor axes. The conduction band energy near a valley
minimum is \begin{equation}
E_{\text{CB}}=\frac{\hbar^2}{2}\left(\frac{k_x^2}{m_x}+\frac{k_y^2}{m_y}
+\frac{k_z^2}{m_z}\right),
\end{equation}
where $m_{x,y,z}$ are constant effective masses for momentum in a given
direction in $R$, and $\vec{k}= k_x\vec{\hat{x}} + k_y\vec{\hat{y}} +
k_z\vec{\hat{z}}$ is the wave vector relative to the subband minimum.

In QCLs, the quantization effective mass $m_q$ describes the variation 
of conduction band potential in the growth direction, \emph{i.e.\/} $m_q
= m_z$. The density-of-states mass $m_d$ accounts for in-plane motion
and is anisotropic with respect to the in-plane wave vector. An
isotropic approximation, $m_d=\sqrt{m_{x}m_{y}}$ is commonly used
however.\cite{JAPSchenk1996, PRBMizuno1993} The effective mass values
are summarized in Tbl.~\ref{tbl:EffMass}.

In the (001) orientation, the major axes of all four $\Delta_{xy}$ 
valleys lie in-plane, so $m_q=m_t$ and $m_d=\sqrt{m_{t}m_l}$.  
Conversely, the major axis of each $\Delta_z$ valley points in the 
growth direction, so $m_q=m_l$, while $m_d$ is isotropic and equal to
$m_t$.

It is slightly inaccurate to treat the $\Delta_z$ subbands as being
twofold degenerate in a simple effective mass approximation (EMA) as 
intervalley mixing splits the degeneracy in quantum confined 
systems.\cite{JPSJOhkawa1977_1}  We have accounted for this in symmetric
systems in a modified EMA and shown that the effect may be large in
narrow quantum wells.\cite{PRBValavanis2007} However, asymmetric 
structures require a computationally expensive atomistic
calculation.\cite{PRBBoykin2004, PRBFriesen2007} As the splitting is
at most a few meV in weakly confined states, the effect was neglected
in the QCL simulations presented in this work.

In (111) heterostructures, the situation is much simpler.  All six 
$\Delta$ valleys have $m_q=0.26$ and identical confining potentials.
Intervalley mixing is absent as each valley lies at a different value of 
$k_{xy}$.  The subbands are therefore sixfold degenerate within the EMA\@.

In summary, the conduction band for (001) heterostructures is
complicated by effective mass anisotropy.  The $\Delta_z$ subbands may 
conceivably be used for QCL design, but the large $m_q$ value severely 
limits the oscillator strength and hence the gain.\cite{Davies1998} The 
$\Delta_{xy}$ subbands have lower $m_q$, but states are weakly 
confined by the small band offsets and have low populations due to their
high energy.

The (111) orientation however, offers both low effective mass and high 
band offset and shows more promise for QCL designs.  The low mass allows
wider wells to be used than for $\Delta_z$ transitions in (001) systems.
QCL designs in the (111) orientation are therefore more tolerant to 
deviations in layer thicknesses caused by growth errors.

\section{\label{scn:ScattMech}Carrier transport model}
Having established the differences in band structure and effective mass 
between (001) and (111) oriented heterostructures, carrier transport
may now be modeled in QCL structures.  Detailed quantum theoretical 
approaches have been used to simulate intersubband optical emissions in 
quantum wells\cite{PRBWaldmuller2004} and carrier transport in limited 
numbers of subbands in QCLs.\cite{PRBLee2002} They are, however, too 
computationally demanding for use as design tools for large multi-level
QCLs.

Reasonably good agreement has been achieved between experimental results 
and Boltzmann or rate equation based models of bound-to-continuum THz
QCLs in III--V systems.\cite{NatureKohler2002, JAPJovanovic2006}  In the
present work, we have therefore determined subband populations 
using a computationally efficient rate equation 
approach.\cite{JAPJovanovic2006} This has been described in more detail
previously,\cite{PRBValavanis2008} although a brief summary follows.

Several intravalley scattering mechanisms are important in Si/SiGe 
systems. Interface roughness scattering was calculated using the
correlated Gaussian roughness model,\cite{RMPAndo1982} modified for
arbitrary interface geometries\cite{PRBValavanis2008} and alloy disorder
scattering was calculated using a point perturbation
model.\cite{PRBQuang2007, PRLMurphy-Armando2006} Ionized impurity
scattering was determined using a Coulombic interaction model as
described by Unuma\cite{JAPUnuma2003} with Thomas-Fermi
screening,\cite{Davies1998} while electron-electron scattering was
treated as a screened Coulombic interaction as described by
Smet.\cite{JAPSmet1996} The intravalley deformation potential scattering
for electron-acoustic phonon interactions was also included.

Intervalley electron-phonon scattering was determined only for the Si-Si
branch of the deformation potential interaction, as the Ge 
fraction in quantum wells is small.  The $f$ processes, which transfer 
electrons to the perpendicular valleys, are faster than $g$ processes
which transfer electrons to the opposite valley, due to the larger 
number of destination states.\cite{PRBCanali1975} Of the $f$ processes,
the $f$-LA (phonon energy $\hbar\omega_0$ =
46.3\,meV\cite{JAPDollfus1997}) and $f$-TO ($\hbar\omega_0$ =
59.1\,meV\cite{JAPDollfus1997}) interactions are rapid zero-order terms
in the scattering model.\cite{PRBMonsef2002} Scattering rates increase
rapidly with transition energy, and saturate above the phonon energy.

The electron transfer rate from initial subband $i$ to final subband
$f$ is the product of the average intersubband scattering rate
$\overline{W}_{if}$ (due to all scattering processes) and the initial 
subband population $N_i$. Although a simple estimate of current density
is proportional to the total electron transfer rate,\cite{JAPIkonic2004} 
an improved model takes account of the spatial separation of electrons.
The current density is therefore
\begin{equation}
\label{eqn:current}
  J = \frac{q_e}{L_p}\sum\limits_i N_i \sum\limits_f
    \overline{W}_{if}\left(\langle{}z\rangle_f - \langle{}z\rangle_i
  \right),
\end{equation}
where $q_e$ is the electronic charge, $L_p$ is the length of a
structural period and $\langle{}z\rangle$ is the expectation value of
the position operator.

The active region gain or absorption for each transition was calculated 
as\cite{JAPJovanovic2006}
\begin{equation}
  \label{eqn:gain}
  G_{if}(\omega)=\frac{e^2\omega\pi}{cn_0\epsilon_0L_p}
   N_i|z_{if}|^2\operatorname{sgn}(E_{if})L(\omega,E_{if}),
\end{equation}
where $n_0$ is the refractive index of Si, $\epsilon_0$ is the
permittivity of free space, and $c$ is the speed of light \emph{in
vacuo}. $E_{if}$ is the energy difference between the subband minima,
$\operatorname{sgn}(\cdot)$ is the signum function, $z_{if}$ is the dipole
matrix element, and $L(\omega,E_{if})$ is the lineshape for the
transition.  The gain spectrum is found by summing Eqn.~\ref{eqn:gain}
over all transitions.

Although linewidth may be obtained directly from our scattering rate
calculations,\cite{JAPUnuma2003} or from more sophisticated
models,\cite{PRBWaldmuller2004} several important implementation 
issues, such as how to treat the extremely broad absorptions into weakly
bound higher energy subbands, are beyond the scope of the present work.
Normalized Lorentzian lineshapes centered about $E_{if}$ have been
observed in THz III-V systems, with full-width at half-maximum around
2\,meV at low temperatures.\cite{NatureKohler2002}  We therefore
calculated gain spectra using linewidths in the range 1.5--2.5\,meV.
At higher lattice temperatures, the increased scattering rates cause
linewidth broadening,\cite{PRBWaldmuller2004, APLPage2001} and
consequently we expect our higher linewidth results to be more realistic
at higher temperatures.

\section{QCL performance}
We have described the advantages of the (111) orientation for Si/SiGe 
QCLs in general terms and have previously predicted net gain in a novel 
phonon depopulation QCL.\cite{APLLever2008} In this section, we present
a bound-to-continuum active region design with a double-metal waveguide
and demonstrate that net gain is achievable up to 90\,K. We also show
that reasonable limitations in growth quality due to interdiffusion do
not present a significant obstacle.

\subsection{Active region design}
\begin{figure}[tb]
  \includegraphics{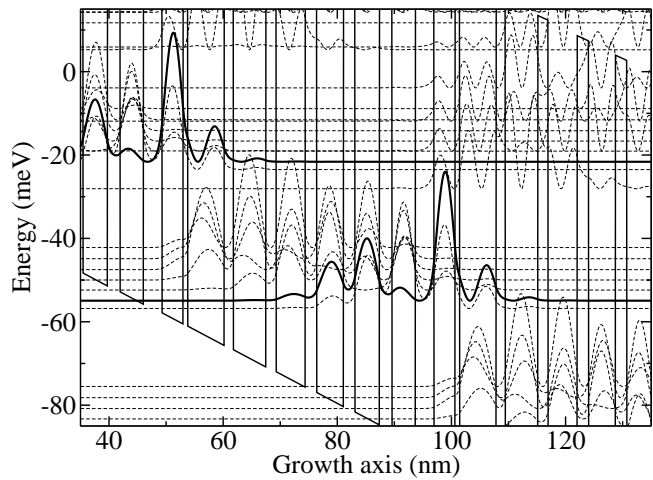}
  \caption{\label{fig:7kV-wavfns}A seven-well bound-to-continuum QCL at
  a bias of 7\ kVcm$^{-1}$ with layer widths of
  \textbf{3.2}/3.7/\textbf{0.8}/6.4/\textbf{1.6}/5.7/
  \textbf{1.8}/5.1/\textbf{2.0}/4.7/\textbf{2.0}/4.3/\textbf{2.2}/4.1,
  where boldface denotes 40\% Ge barriers and lightface denotes pure Si
  wells. Dopants are spread evenly through the structure with a
  concentration of $5\times10^{16}$\ cm$^{-3}$. The conduction band
  potential (solid line) is shown, with spatially dependent probability
  densities superimposed at each subband minimum. The upper laser
  subband is shown in bold, while other subbands are shown as dashed
  lines.}
\end{figure}

\begin{figure}[tb]
  \includegraphics{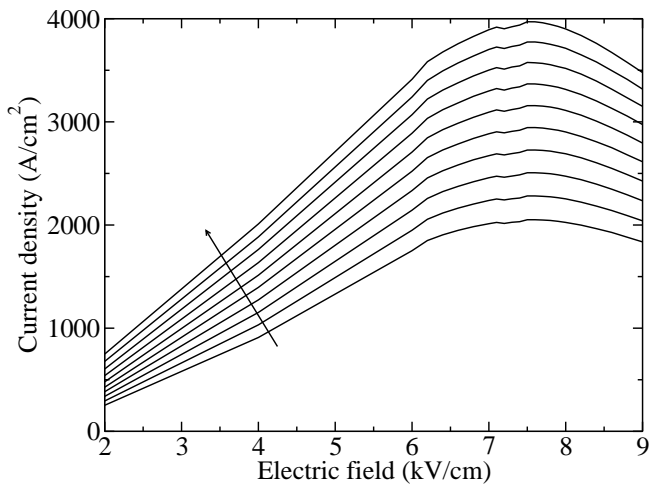}
  \caption{\label{fig:FJ-T4to100}Current density as a function of
  applied electric field for the QCL design in Fig.~\ref{fig:7kV-wavfns}.
  Results are shown for linearly increasing lattice temperatures, in the
  direction of the arrow, between 4 and 100\ K.}
\end{figure}

The band structure in a seven-well bound-to-continuum QCL was calculated
using a single band EMA and is shown in 
Fig.~\ref{fig:7kV-wavfns} for a 7\ kVcm$^{-1}$ applied electric field.
Non-radiative depopulation of the upper laser subband was reduced by
limiting the higher energy scattering processes.  Pure Si wells were 
used to minimize alloy disorder scattering, and a relatively low Ge 
composition of 40\% was selected for the barriers to reduce interface 
roughness scattering. The optical transition energy was chosen to be
significantly smaller than 46.3\,meV to reduce $f$-LA and $f$-TO phonon
emission rates.

As modulation doping of donors in Si/SiGe heterostructures may be 
difficult,\cite{SurfSciZhang2006} dopants were assumed to be spread 
evenly throughout the structure.  It was also assumed that all donors 
were ionized at low temperatures.  A donor concentration of 
$5\times10^{16}$\ cm$^{-3}$ (sheet doping density of $2.4\times10^{11}$\ 
cm$^{-2}$) caused negligible internal electric fields, while still
allowing rapid depopulation of upper miniband states by Coulombic
scattering.

\begin{figure}[tb]
  \includegraphics{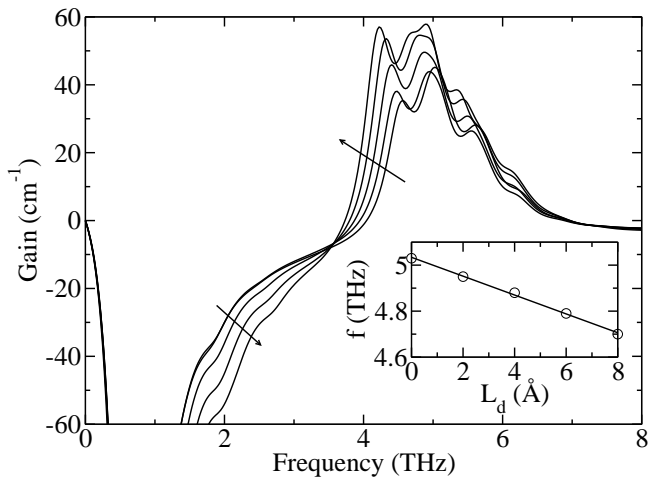}
  \caption{\label{fig:gain7kV}Active region gain at 7\ kVcm$^{-1}$, and
  4\ K lattice temperature as a function of THz frequency for the design
  in Fig.~\ref{fig:7kV-wavfns}. Arrows indicate linearly increasing
  diffusion lengths from 0 to 8\ \AA\@. The resulting frequency shift of
  the largest gain peak is shown inset.}
\end{figure}

Figure~\ref{fig:gain7kV} shows gain at frequencies around 5\,THz,
due to transitions from the upper laser subband to upper miniband 
states, for 4\,K lattice temperature and 7\,kVcm$^{-1}$ applied electric
field. Figure~\ref{fig:FJ-T4to100} shows that a current density of
2\,kAcm$^{-1}$ corresponds to these conditions, and doubles as lattice 
temperature increases to 100\,K. An energy balance 
approach\cite{JAPJovanovic2006} was used to find electron temperatures. 
Assuming an identical temperature in each subband, we found that 
electron temperatures increased from 110 to 150\,K as lattice 
temperatures increased from 4 to 100\,K. 

\begin{figure}[tb]
  \includegraphics{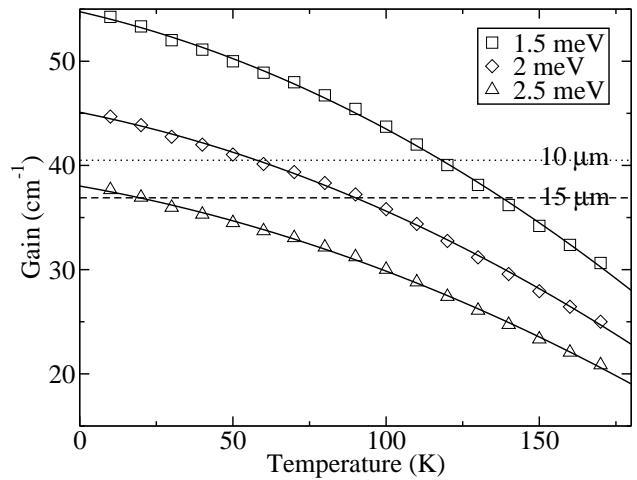}
  \caption{\label{fig:GmaxT}Peak gain near 5\,THz as a function of
  lattice temperature for the QCL design in Fig.~\ref{fig:7kV-wavfns}.
  Results are given for linewidths of 1.5, 2.0 and 2.5\,meV. The
  threshold gains for 10 and 15\,$\mu$m active region thicknesses are
  shown as dotted and dashed lines respectively.}
\end{figure}

Figure~\ref{fig:GmaxT} shows the peak gain near 5\,THz as a function of
lattice temperature for 1.5, 2.0 and 2.5\,meV transition linewidths.  As
discussed in section \ref{scn:ScattMech}, linewidth increases with
temperature, and we expect the low-temperature gain to be enhanced and
the high-temperature gain to be limited accordingly.

To provide a fair comparison with $\Delta_z$ subbands in the (001)
orientation, an equivalent (001) design was developed. Firstly, the
$\Delta_z$ conduction band offset was matched to that of the (111)
design by setting the barrier Ge fraction to 26\%. Noting that the
subband energies are much smaller than the barrier energy, an infinite
quantum well approximation was used to transform the layer widths and
preserve subband separation,
\begin{equation}
  l_k^{(001)} = l_k^{(111)}\sqrt{\frac{m_q^{(111)}}{m_q^{(001)}}}=0.533 l_k^{(111)}.
\end{equation}
This scaling also preserves the barrier transparency, \emph{i.e.} the
coupling between adjacent wells. Finally, the applied electric field 
was increased to 13.1\,kVcm$^{-1}$ to account for the reduced period length.

The $\Delta_z$ subbands in the (001) system were at sufficiently low
energy compared with the $\Delta_{xy}$ valleys, for $f$-phonon 
interactions to be negligible. With the larger electric field, and with
only slow $g$-phonon interactions available to cool the electron
distribution, the steady-state electron temperature was greatly
increased to 290\,K for a lattice temperature of 4\,K.  Consequently,
thermal backscattering led to a much lower peak gain coefficient of 
0.2\,cm$^{-1}$ at 6\,THz, corresponding to a current density of 
1.1\,kAcm$^{-1}$.  Growth of such a structure is also expected to be
challenging, as the minimum layer thickness was reduced to 4\,\AA{}.

\subsection{Waveguide design}
A suitable waveguide was designed using a one-dimensional simulation.
The propagation constant was obtained using a transfer matrix
method\cite{JLTAnemogiannis1999} and the complex permittivities were
found using the bulk Drude model using Si/SiGe material parameters from
Ref.~\onlinecite{phd:Palankovski2000}.  The active region was modelled
as bulk Si$_{0.9}$Ge$_{0.1}$ (to match the virtual substrate), with a
doping concentration of $5\times{}10^{16}$\,cm$^{-3}$.

Surface-plasmon configurations\cite{NatureKohler2002} were found to be
unsuitable, due to the low confinement factor $\Gamma$ and large
waveguide losses $\alpha_{\text{w}}$.  We therefore chose a metal-metal
configuration, which has proved successful in GaAs-based THz
QCLs.\cite{APLWilliams2003}  The active region was enclosed between a
pair of highly doped ($n=10^{19}$\,cm$^{-3}$), 20\,nm thick Si layers,
followed by the metallic layers. The optical properties of the metallic
layers are given in Ref.~\onlinecite{AOOrdal1985}.

An initial design, using gold metallic layers and a 10\,$\mu$m
thick active region gave $\alpha_{\text{w}}$=50.7\,cm$^{-1}$ and
$\Gamma$=0.99. Assuming mirror losses of $\alpha_{\text{m}}$ =
1\,cm$^{-1}$,\cite{JAPKohen2005} the threshold gain was determined as
$g_{\text{Th}}=(\alpha_{\text{w}} + \alpha_{\text{m}})/\Gamma =
51.7$\,cm$^{-1}$, which was too high to achieve lasing.

The highest temperature operation of a GaAs-based THz QCL was
achieved recently by using copper instead of gold
layers.\cite{OptExpBelkin2008} Incorporating this into our waveguide
reduces the threshold gain to 40.5\,cm$^{-1}$.  As shown in 
Fig.~\ref{fig:GmaxT}, this permits lasing up to T=58\,K for a
2\,meV linewidth. By increasing the active region thickness to
15\,$\mu$m, the threshold gain was reduced further to 36.9\,cm$^{-1}$,
which permits lasing up to T=90\,K.

\subsection{Growth variations}
We restricted our (111) design to layer thicknesses above 8\,\AA{} as
Si/SiGe epitaxy is not as well established as in III-V systems. The
thinnest barrier would ideally be thinner than this to increase the
dipole matrix element between the bound subband and the upper miniband
states. The requirement for thin barriers is less important in (111)
heterostructures than in (001) however, as the quantization effective
mass is smaller and the matrix element is larger.

Ge surface segregation has been observed in (001)
heterostructures,\cite{SurfSciZhang2006} and presumably this will also
be the case in (111) systems. The geometry of the thinnest layers in a
QCL is expected to change considerably as a result. We have previously
shown however, that a limited amount of interdiffusion is tolerable,
although changes in transition energies are
expected.\cite{PRBValavanis2008}

Figure~\ref{fig:gain7kV} shows that gain increases slightly as a
function of interdiffusion length, $L_d$ up to 8\,\AA{} for our QCL
design. This is due to the thinnest barrier being degraded, increasing 
the dipole matrix element for optical transitions. The inset in the 
figure shows that the center of the largest gain peak correspondingly
shifts from around 5 to 4.7\,THz, as the upper laser subband energy
decreases.\cite{PRBValavanis2008}

\section{Conclusion}
We have shown that intersubband lasing in the $\Delta$ valleys of 
Si/SiGe heterostructures becomes viable in the (111) orientation.
Although the $\Delta_z$ conduction band offset in the (001) orientation
is large, the \emph{usable\/} energy range was shown to be superior in
the (111) orientation due to the sixfold valley degeneracy. The
quantization effective mass was also shown to be much smaller in (111)
heterostructures, and complications due to intervalley mixing and
uniaxial strain splitting are avoided.

We have presented a bound-to-continuum design for a (111) Si/SiGe QCL
and investigated several options for waveguides.  A surface-plasmon 
waveguide was shown to be inadequate, while good results were achieved 
for a double-metal configuration using copper metallic layers. We have
shown using a self-consistent rate-equation/energy balance calculation
that net gain at 5\,THz is possible up to a lattice temperature of
90\,K, with a low-temperature current density of 2\,kAcm$^{-1}$ for a
15\,$\mu$m thick active region.

The (111) design was found to be vastly superior to a (001) oriented
equivalent, due to the phonon-mediated electron cooling and the reduced 
effective mass. It was also shown to be tolerant to, and indeed
improve slightly, with Ge interdiffusion lengths up to 8\,\AA\@.

\begin{acknowledgments}
  This work is supported by EPSRC Doctoral Training Allowance funding
  and DTI-MNT contract 491: ``Fast THz Cameras''.
\end{acknowledgments}
\bibliography{nSiGePaper}
\end{document}